# Prediction of Pneumonia and COVID-19 Using Deep Neural Networks

Md Sabbirul Haque
IEEE Professional Community
IEEE
sabbir465@gmail.com

Md Siam Taluckder
Phillip M. Drayer Department of Electrical Engineering
Lamar University
mtaluckder@lamar.edu

Siam Bin Shawkat
Computer Science Department
American International University - Bangladesh
sb.shawkat@gmail.com

Md Ahnaf Shahriyar
College of Technology and Engineering
Westcliff University
Irvine, CA
m.shahriyar.229@westcliff.edu

Md Abu Sayed
Department of Professional Security Studies.
New Jersey City University
Jersey City, New Jersey.
msayed@njcu.edu

Chinmoy Modak
Department of Computer Science
Florida Polytechnic University
Lakeland, Florida
modakc64@gmail.com

*Abstract*- **Pneumonia, caused by bacteria and viruses, is a rapidly spreading viral infection with global implications. Prompt identification of infected individuals is crucial for containing its transmission. This study explores the potential of medical image analysis to address this challenge. We propose machine-learning techniques for predicting Pneumonia from chest X-ray images. Chest X-ray imaging is vital for Pneumonia diagnosis due to its accessibility and cost-effectiveness. However, interpreting X-rays for Pneumonia detection can be complex, as radiographic features can overlap with other respiratory conditions. We evaluate the performance of different machine learning models, including DenseNet121, Inception Resnet-v2, Inception Resnet-v3, Resnet50, and Xception, using chest X-ray images of pneumonia patients. Performance measures and confusion matrices are employed to assess and compare the models. The findings reveal that DenseNet121 outperforms other models, achieving an accuracy rate of 99.58%. This study underscores the significance of machine learning in the accurate detection of Pneumonia, leveraging chest X-ray images. Our study offers insights into the potential of technology to mitigate the spread of pneumonia through precise diagnostics.**

## I. Introduction

Pneumonia is a respiratory infection characterized by the infiltration of air cells within one or both lungs, caused by viral, bacterial, or dual agents. This infection leads to symptoms such as a productive cough accompanied by phlegm or pus, elevated body temperature (fever), sensations of coldness (chills), and respiratory distress due to the accumulation of fluid or pus within the air cells. Pneumonia is brought about by a variety of microorganisms, encompassing fungi, viruses, and bacteria.

The severity of pneumonia ranges from mild forms to instances of life-threatening conditions. Individuals aged over 65, individuals with pre-existing health conditions, and those possessing compromised immune systems are particularly susceptible to severe manifestations of pneumonia. In contrast to bacterial pneumonia, which often follows a deterioration pattern prior to recovery, viral pneumonia frequently undergoes spontaneous resolution. Pneumonia can manifest in singular or dual lung involvement, dependent on the extent of the infection and its location within the respiratory system [1].

The manifestation of pneumonia can vary from being sufficiently severe to necessitate hospitalization, or so mild that its presence goes almost unnoticed. Among the different types of pneumonia, bacterial pneumonia is the most prevalent and typically presents with more pronounced symptoms compared to other forms. Such symptoms often demand immediate medical intervention. Bacterial pneumonia can elicit a sudden or gradual onset of adverse effects. These effects may encompass a potentially hazardous fever, reaching temperatures as high as 105 degrees Fahrenheit, accompanied by a rapid elevation in breathing and heart rates, coupled with profuse sweating. Additionally, reduced oxygen levels in the blood can lead to a bluish discoloration of the lips and nailbeds. The cognitive state of a patient might exhibit confusion or irrationality. In the case of viral pneumonia, symptoms typically manifest over a span of several days. Within a day or two, these symptoms generally exacerbate, incorporating an intensified cough, difficulty in breathing, and increased muscular pain. High fever and bluish discoloration of the lips can also manifest as potential side effects.

Scientists have come up with a new and affordable way to detect pneumonia, which is a lung infection. They suggest using X-ray images of the chest to look for signs of pneumonia [1,2,3]. pneumonia resembles viral pneumonia in appearance on radiographs, its radiological detection can be challenging [1, 4]. With the high number of suspected cases daily, distinguishing between the two may not always be possible due to the need for specific expertise. Automation and machine learning have been used by researchers to close this gap. The purpose of this research is to examine how machine learning algorithms can be used to identify pneumonia in chest X-ray images. We discuss the performances of various machine learning algorithms and their implications in diagnosing pneumonia with limited resources. The findings from this study can be greatly beneficial for the development of precise and cost-effective methods for detecting pneumonia. We demonstrate that the use of our proposed machine learning algorithms in radiological imaging can significantly improve the accuracy of detecting pneumonia.

II. Literature Review

Rajasenbagam et al [5]. In this research, it was suggested that chest X-ray images could be used to detect pneumonia infection in the lung using a Deep Convolutional Neural Network. Pneumonia Chest X-ray Dataset: 12,000 images of infected and uninfected chest X-rays were used to train the proposed Deep CNN models. The augmented images were made by utilizing the fundamental manipulation strategies and the Deep Convolutional Generative Adversarial Network (DCGAN). The proposed Deep CNN model was developed using the VGG19 network. In the unseen chest X-ray images, the proposed Deep CNN model had a classification accuracy of 99.34%. The proposed deep CNN's performance was compared to that of cutting-edge transfer learning methods like AlexNet, VGG16Net, and InceptionNet. The comparison results demonstrate that the proposed Deep CNN model outperformed the other methods in terms of classification performance.

K. S. and D. Radha et al [6]. The proposed work is to use X-rays, one of the medical imaging techniques used to examine the health of patients with lung inflammation, to identify COVID-19 and pneumonia patients. For the identified dataset, the appropriate Convolutional Neural Network Model is chosen. On the actual dataset of lung X-ray images, the model finds patients with pneumonia and COVID-19. Images are trained for Normal, COVID-19, and Pneumonia classifications after being pre-processed. The disease is detected by selecting the appropriate features from each dataset's images following pre-processing. The result indicates that COVID and pneumonia were accurately identified. COVID versus Normal is more accurate than COVID versus Pneumonia among these two options. With 80% and 91.46%, respectively, this method identifies not only COVID or pneumonia but also its subtypes—bacterial pneumonia and viral pneumonia. Utilizing the proposed model, COVID, bacterial pneumonia, and viral pneumonia can all be identified and distinguished quickly, making it easier to implement prompt and appropriate solutions.

D. Varshni et al [7]. In this study, Pre-trained CNN models used as feature extractors and various classifiers for the classification of abnormal and normal chest X-rays are evaluated. The most suitable CNN model is selected analytically for this purpose. Pre-trained CNN models and supervised classifier algorithms can be very helpful in the analysis of chest X-ray images, particularly for the purpose of detecting pneumonia, as demonstrated by the statistical results obtained.

Kareem et al [8]. This paper surveys and examines the use of computer-aided methods in the detection of pneumonia. It also suggests a hybrid model that uses real-time medical image data in a privacy-preserving manner to effectively detect pneumonia. In this paper, we'll look at how X-rays and other preprocessing methods can identify and classify a wide range of diseases. The survey also looks at how different machine learning technologies like k-nearest neighbor (KNN), convolution neural network (CNN), RESNET, Chex Net, DECNET, and ANN can be used to find pneumonia. An extensive literature review is conducted within this article to ascertain the possibilities of amalgamating hospital and medical institution datasets in order to enhance the training of machine learning models for improved disease detection accuracy and efficiency,

A. Pant et al [9]. The main motivation behind this research was utilizing the X-ray images of the patients alone to identify pneumonia. Since doctors need to run a lot of specific tests to figure out if a patient has pneumonia. To address this challenge, the researchers devised an ensemble of two deep-learning models. This ensemble approach serves to streamline doctors' tasks, achieving notable accuracy when tested against both unseen data and relevant research papers. This study culminates in the proposition of an ensemble deep learning model, presenting a comprehensive solution to the issue at hand.

M. Yaseliani et al [10]. In this research, using three classification methods, a novel hybrid Convolutional Neural Network (CNN) model is proposed. CXR images are classified using Fully Connected (FC) layers in the first classification method. The weights with the highest classification accuracy are saved after this model is trained for several epochs. The most representative CXR image features are extracted using the second classification method's trained optimized weights, and Machine Learning (ML) classifiers are used to classify the images. CXR images are classified using an ensemble of the proposed classifiers in the third classification method. With an accuracy of 98.55%, the results indicate that the proposed ensemble classifier employing Support Vector Machine (SVM), Radial Basis Function (RBF), and Logistic Regression (LR) classifiers performs the best. In the end, this model is used to make a web-based CAD system that can help radiologists find pneumonia with a lot of accuracy.

III. Methodology

In this section, we discuss various approaches based on machine learning algorithms that we used and trained for detecting pneumonia. Machine learning models like Inception Resnet-V2, Inception-v3, Resnet-50, Xception, and Densenet-121 are used in our prediction. All the models give us good results but comparatively one model gives us the best result for the prediction of pneumonia. So, below we discuss the methodologies of various models used in this study.

a) Inception-ResNet-V2

Convolutional neural architecture Inception-ResNet-v2 adds residual connections to the Inception family of architectures. The residual Inception Block is Inception-ResNet-V2's fundamental building block. Before the addition to match the depth of the input, a 1 × 1 convolution filter expansion layer is used after each block to scale up the dimensionality of the filter bank. Batch normalization is only used on top of the traditional layers in this architecture. The Inception-ResNet-V2 image input size is 299. It is 164 layers deep. Multiple-sized convolutional filters with residual connections are incorporated into the Residual Inception Block [13,16]. This architecture reduces training time and avoids the degradation issue associated with deep networks by making use of residual connections. Our refined Inception-ResNet-V2 model for COVID-19 and pneumonia classification is shown in Figure 1.

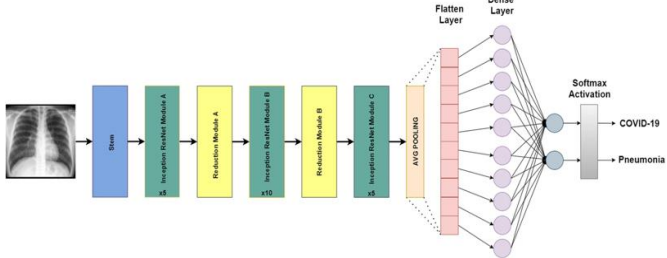

Fig 1: Inception-ResNet-V2 design for binary Classification

### b) Inception-V3

This paragraph explains the Inception-V3 model used in the study. The model used in this study comprises there are 11 inception modules spread across 484 layers in InceptionV3. It has a picture input size of 299 × 299. Convolution filters, pooling layers, and the ReLu activation function make up each module [1,19]. By factorizing convolutions, InceptionV3 reduces the number of parameters without affecting the efficiency of the network. In addition, InceptionV3 made a novel proposal for reducing the number of features. Our refined model of InceptionV3 for COVID-19 and pneumonia classification is shown in Figure 2.

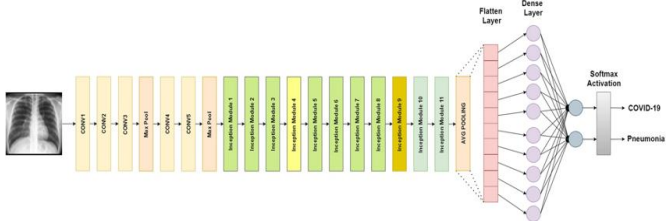

Fig 2: Inception-V3 design for binary Classification

An equivalent quantity of samples was randomly selected for both training and validation, constituting a pneumonia dataset. Within this sample pool, 20% of the cases were allocated for validation purposes, leaving the remaining 80% for training endeavors. To enhance testing results and prevent the issue of overfitting concerning pneumonia instances, the remaining samples were held back. Irrespective of the cases present in the test dataset, the training and validation tests were executed in a balanced manner to achieve optimal accuracy. This approach was adopted due to previous studies showcasing the necessity of a well-balanced training dataset to attain precise outcomes.

### c) Dense-Net-121

The DenseNet121 model can only be used with an RGB image that is at least 224 by 224 pixels in size. The model's 121 layers are made up of more than 8 million parameters. Dense Blocks, the core of DenseNet121, change the number of filters while maintaining the dimensions of the feature map. In between the blocks, the Transition layers use batch normalization for downsampling. The final fully linked layer with Soft Max activation is replaced in this study by a custom classifier, as depicted in Figure 3.

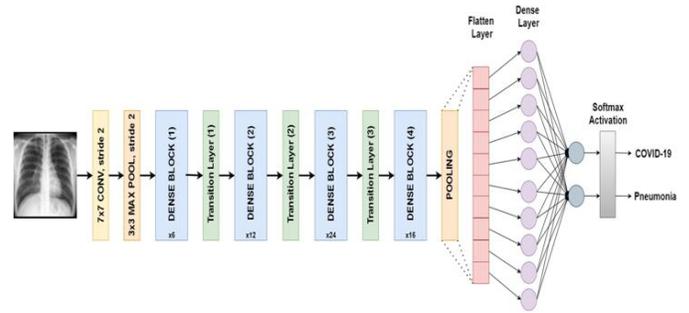

Fig 3: Dense Net 121 for binary Classification

### d) Res-Net-50

The ResNet50 design for residual networks consists of 48 convolutional layers, one Max-Pool layer, and one Average Pool layer. Each convolution block in ResNet50 has three convolutional layers. ResNet50 has one identification block and three convolutional layers. The model is capable of being trained with over 23 million distinct parameters. In this study, the ResNet50 model was modified to take into account pneumonia categorization and COVID-19. This is depicted in Figure 4.

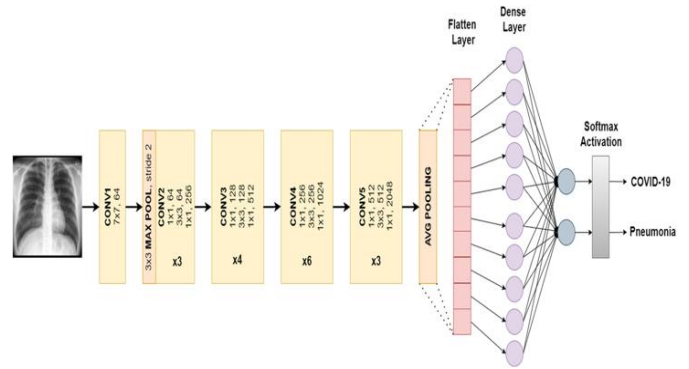

Fig 4: Res-Net-50 for binary Classification

### e) Xception

In 2016, Chollet, the creator of the Keras library, proposed a machine-learning model named Xception. It is an adaptation of the Inception architectures in which depth-wise separable convolutions take the place of the Inception modules. On the ImageNet dataset, Xception performed better than the conventional InceptionV3, attaining higher Top-1 and Top-5 accuracy. Xception has roughly the same number of parameters as InceptionV3 (around 23 million). Our refined Xception model for COVID-19 and pneumonia classification is shown in Figure 5.

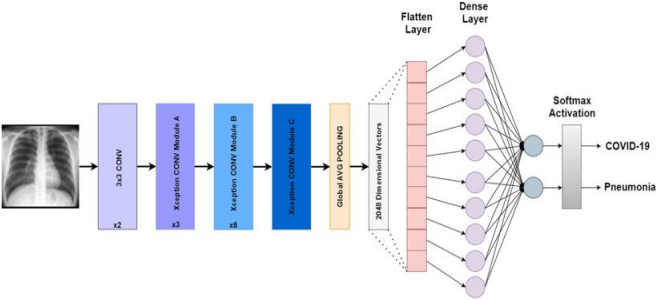

Fig 5: Res-Net-50 for binary Classification

**Model Training and validation**

This research employed five distinct machine learning models (namely, Inception-ResNet-v2, Inception-v3, DenseNet121, ResNet50, and Xception). To ensure uniformity, all images within the dataset underwent resizing to dimensions of 224 × 224 pixels. The algorithm's architecture was fashioned, and the convolutional neural network (CNN) was implemented using Tensor Flow 2.4 and the Keras API. Training for these models took place on a 12 GB NVIDIA Tesla K80 GPU. Throughout the training phase, model performance was assessed by their capacity to accurately predict probabilities aligned with the ground truth, utilizing the categorical cross-entropy loss function.

In this investigation, we amassed a compilation of images from various openly accessible databases. A total of 3525 pneumonia photographs and 1123 images depicting COVID-19 were procured to serve as the foundation for training, validation, and testing of the models. The images underwent a uniform standardization procedure, involving resizing to dimensions of 224 by 224 pixels, irrespective of their initial size variations. Within the COVID-19 samples, a random selection of 10% was earmarked for testing purposes, while the remaining images were partitioned, with 80% designated for training and the remaining 20% allocated for validation.

## IV. Result and Discussion

Table 1, Chart 1, and Figures 6, 7, 8, 9, and 10 provide visual representations of the accuracy and loss metrics for each refined model throughout the stages of both training and validation. The table presents the presentation measurements of different models on a characterization task. DenseNet-121, ResNet-50, Xception, and Inception ResNet-V2 are the models being evaluated. Accuracy, precision, recall, and the F1 score are some of the measured metrics.

DenseNet-121 has the highest accuracy of 99.48 percent when looking at the values for accuracy, indicating that it can correctly classify instances. With an accuracy of 99.26 percent, ResNet-50 closely follows. Both Inception-V3 and Inception ResNet-V2 have high levels of accuracy, reaching 98.96 percent and 98.21 percent, respectively. On the other hand, Xception has a slightly lower accuracy of 98.34 percent.

A measure of precision is the proportion of correctly predicted instances. Again, DenseNet-121 does well in this area, getting a precision score of 98.72 percent. With a precision of 99.42 percent, ResNet-50 closely follows, demonstrating its capacity to reduce false positives. The precision levels of Inception-V3 and Xception are, respectively, 98.67% and 98.00%. Among the models, Inception ResNet-V2 has the lowest precision but maintains a respectable score of 98.10 percent.

Table 1: Model Performance

| Model | Accuracy (%) | Precision (%) | Recall (%) | F1 Score (%) |
|---|---|---|---|---|
| Inception ResNet-V2 | 98.21 | 98.10 | 97.06 | 98.06 |
| Inception-V3 | 98.96 | 98.00 | 98.16 | 97.12 |
| Dense-Net-121 | 99.48 | 98.72 | 99.38 | 99.42 |
| Res-Net-50 | 99.26 | 99.42 | 99.31 | 99.30 |
| Xception | 98.34 | 98.67 | 99.01 | 98.89 |

The model's ability to correctly identify positive instances is measured by the recall, also known as sensitivity or true positive rate. With a score of 99.38 percent, DenseNet-121 excels at recall. ResNet-50 is next, with a recall of 99.31 percent. With a recall value of 99.01 percent, Xception outperforms the other models. The recall rates of Inception-V3 and Inception ResNet-V2 are respectively 98.16% and 97.06%.

The F1 score provides a balanced evaluation of the model's performance because it combines precision and recall. DenseNet-121 has the highest F1 score, 99.42 percent, indicating that it performs classification tasks well overall. With an F1 score of 99.30 percent, ResNet-50 comes in close behind. F1 scores of 97.12 percent and 98.89 percent are displayed by Inception-V3 and Xception, respectively. The F1 score of Inception ResNet-V2 is 98.06%.

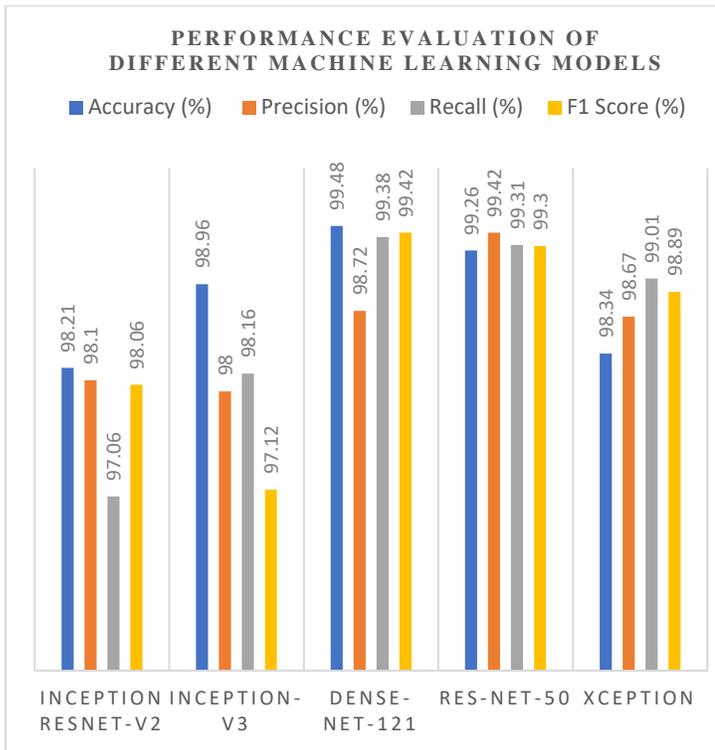

Chart 1:
Assessment of the Effectiveness of Varied Machine Learning Models.

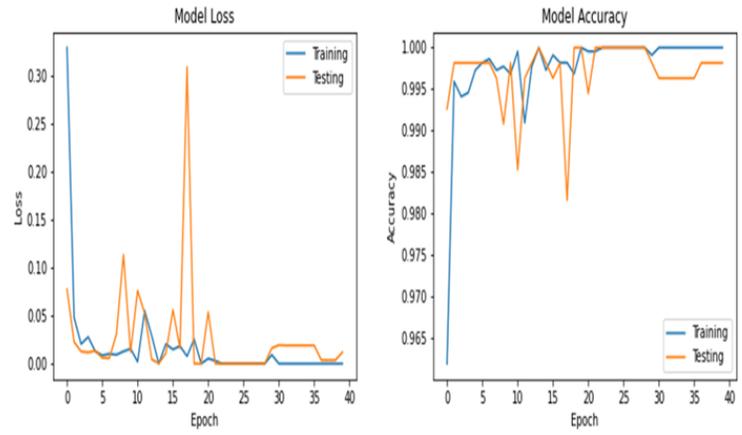

Fig 8: Graph depicting the accuracy of the DenseNet model.

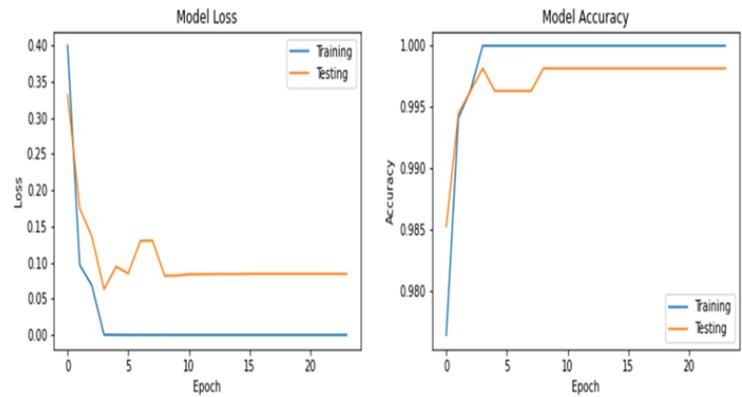

Fig 9: Graph depicting the accuracy of the ResNet-50 model.

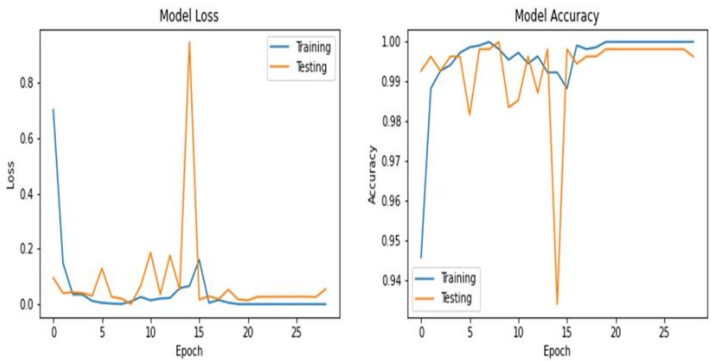

Fig 6: Graph illustrating the accuracy of the Inception-ResNet-V2 model.

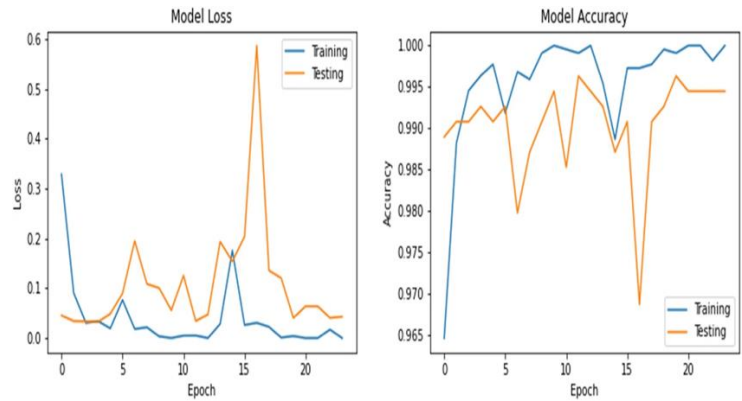

Fig 10: Xception model accuracy graph

V. Conclusion and Future Work

In our study, we consider various advanced machine learning algorithms and compared their performances. Among all models, DenseNet-121 achieves the highest rankings in terms of accuracy, precision, recall, and F1 score. In most metrics, ResNet-50 performs similarly well to DenseNet-121. In contrast, Inception ResNet-V2 receives slightly lower scores compared to the robust performance of Inception-V3 and Xception. Overall, the prowess of machine learning in healthcare is epitomized by DenseNet-121's capacity to accurately identify pneumonia through chest X-rays. This underscores the significance of ongoing research and innovation in this realm. While further research is necessary to validate DenseNet-121's efficacy on a larger scale, the initial findings are highly promising. The potential of DenseNet-121 to revolutionize pneumonia detection through chest X-rays could significantly impact the diagnosis and treatment of this disease, particularly in regions with limited access to medical resources and testing.

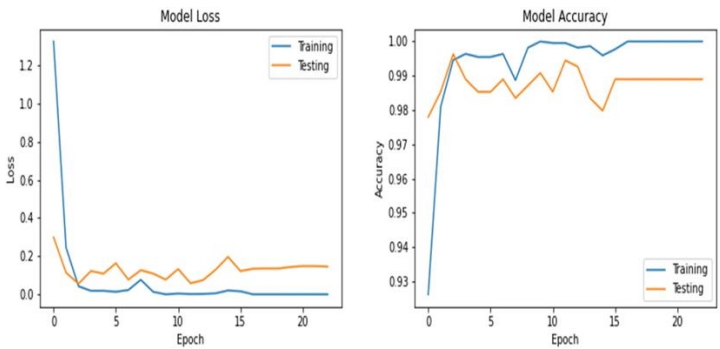

Fig 7:
Graph depicting the accuracy of the Inception-V3 model.